\documentclass{epl}

\title{Aging dynamics and density relaxation in kinetic lattice 
gases under gravity
}
\shorttitle{Aging dynamics of lattice gases}
\author{Y. Levin\inst{1}\thanks{E-mail: \email{levin@if.ufrgs.br}} 
   \and J. J. Arenzon\inst{1}\thanks{E-mail: \email{arenzon@if.ufrgs.br}}
   \and M. Sellitto\inst{2}\thanks{E-mail: \email{msellitt@ens-lyon.fr}}}
\institute{\inst{1} Instituto de F\'{\i}sica, 
  Universidade Federal do Rio Grande do Sul \\ 
  Caixa Postal 15051, CEP 91501-970, Porto Alegre, RS, Brazil \\
  \inst{2} Laboratoire de Physique, \'Ecole Normale Sup\'erieure de Lyon
  - 46 All\'ee d'Italie, F-69364 Lyon cedex 07, France}         

\pacs{05.40.-a}{...}
\pacs{05.70.Fh}{...}

\newcommand{\rhor}{\rho_{\scriptscriptstyle\rm R}}
\newcommand{\kb}{k_{\scriptscriptstyle\rm B}}

\begin{document}

\maketitle

\begin{abstract}
We present an analytical approach to the out of equilibrium dynamics 
of a class of kinetic lattice gases under gravity.
The location of the jamming transition, 
the critical exponents, and 
the scaling functions characterizing the relaxation processes are determined.
In particular, we find that logarithmic compaction and simple aging  
are intimately related to the Vogel-Fulcher law, while power-law 
compaction and super-aging behavior occur in  presence of a
power-law diffusion.
\end{abstract}





Granular materials set in rapid motion by vibration~\cite{Campbell} 
exhibit features such as universal velocity distribution~\cite{RoMe}, 
and ``de Gennes narrowing'' --- a physical signature of the cage 
effect~\cite{WaHa} --- which in spite of the non-equilibrium nature of the 
stationary state, closely resemble those observed in simple liquids at 
thermal equilibrium~\cite{HaDo}.
In the opposite, quasi-static flow limit, slow compaction phenomena
appear~\cite{Chicago}.
During compaction, the free volume available to grains
decreases, and the mobility steeply 
falls to zero, hence aging phenomena are expected to 
occur~\cite{Struik,Bouchaud,BoCuKuMe}, as is confirmed in several numerical 
simulations~\cite{NiCo,BaLo,TaTaVi,Bideau}.
It has been suggested that in this regime a granular 
material should resemble a highly viscous liquid or a 
glass~\cite{Sam}, and several approaches have 
been proposed to describe different aspects of the granular 
compaction dynamics.
These are mainly based on Langevin~\cite{Anita} and 
Fokker-Planck equation~\cite{EdGr}, 
fluctuating nonlinear hydrodynamics~\cite{Mazenko}, 
and mode-coupling theory~\cite{Cugliandolo}.

Two effects are responsible for the unusual behavior of a compacting 
granular material.  
First, collisions between the particles are inelastic, and energy has 
to be constantly pumped into the system.  
Second, at high packing density, steric hindrance, and the associated cage 
effect, play  a crucial role very similar to the one observed in amorphous 
systems. 
In this letter we shall be concerned precisely with this second effect, 
which in the case of gentle shaking is the dominant one.
A number of lattice-gas  models have been introduced to study numerically 
the slow dynamics induced by this effect~\cite{kob93,Tetris,Anita1}.  
Our approach allows to characterize in a precise way the density 
relaxation and the aging dynamics in terms of the particle mobility. 
We show the existence of two relaxation regimes (exponential 
and aging) separated by a jamming transition.
Using scaling arguments we investigate the asymptotic long-time 
behavior of the density and of the mean-square displacement,  
and compare the results with the numerical solution of the
diffusion equation.
We also check our analytic results with the Monte Carlo simulation
of a gravity-driven version of the  Kob-Andersen model 
(KA)~\cite{kob93,SeAr00}. In this model, 
non-interacting particles are allowed to move
only if a local kinetic constraint 
on the occupation of the nearest neighboring sites is
satisfied.


\section{The model}

Consider a box of height $H$ 
in contact with a particle reservoir of density $\rhor$
located at $z=H$. All lengths are measured in units of the particle diameter.
A constant gravitational field of strength $g$ acts in the $-z$ direction,
forcing particles from the reservoir to enter into the box.  
The dynamical evolution of the local particle density $\rho(z,t)$
is governed  by the continuity equation, $\partial_t \rho(z,t)
+\partial_z J(z,t) = 0$, with current given by the Fick's law,  
$J(z,t)=-\Gamma(\rho) \partial_z \mu(z,t) $. $\Gamma(\rho)$
is the Onsager mobility and $ \mu(z,t)= \frac{\delta F}{\delta \rho}$
is the local chemical potential.
For simplicity we shall suppose that the only interaction between
the grains is due to the hard core repulsion. 
The exact Helmholtz free energy functional~\cite{lev00}
for the lattice gas version of this model is
\begin{equation}
\label{2}
F[\rho(z,t)]= \kb T \int_0^H \!\!dz\left[
(1-\rho)\ln(1-\rho)+ 
\rho\ln \rho
+\gamma \, z \, \rho \right] \,,
\end{equation}
where  $\gamma= m g/\kb T $ is the inverse gravitational length.
In the case of granular materials the thermal energy of the grains is 
negligible so that $T$ is a function of the externally imposed
vibration intensity~\cite{hay97}.  It is 
neither the physical temperature nor the ``granular temperature'' 
usually associated with the average kinetic energy~\cite{Campbell}. 
For the KA model, the mobility $\Gamma(\rho)$ vanishes 
as $\Gamma(\rho)= \Gamma_0 \, \rho \, (1-\rho/\rho_{\rm c})^\phi$,
with the critical  threshold $\rho_{\rm c}=0.88$ for the simple cubic 
lattice~\cite{kob93}, and $\rho_c=0.81$ for the BCC lattice. 
In both cases $\phi=3.1$. A similar expression is found for the mobility
of highly packed hard-sphere systems~\cite{got92,tok95}. 
One should note, however, that the free energy functional given by 
Eq.~(\ref{2}) is exact only for the lattice gas.  In the case of 
hard spheres it can be replaced by an approximate expression derived from
the Carnahan-Starling equation of state~\cite{car69,lev00}, without any 
qualitative modifications to the theory. Substituting
$\Gamma(\rho)$ into the continuity equation we are lead to,

\begin{equation}
\label{3}
\frac{\partial \rho(z,t)}{\partial t}=\frac{\partial}{\partial z}
\left\{ \left( 1-\frac{\rho}{\rho_{\rm c}} \right)^\phi
\left[\frac{1}{1-\rho}\frac{\partial \rho}{\partial z} + \gamma \, \rho
\right]\right\}\;,
\end{equation}
where the time is now measured in units of $1/\Gamma_0 \kb T$. 
We should stress that the local density approximation in which
the functional form of mobility is assumed to remain valid
in an inhomogeneous system is highly non-trivial.
One of the goals of this paper is to assess to what extent
this approximation works in the case of kinetic lattice-gas models.
  
\begin{figure}[h]
\onefigure[width=5.5cm, angle=270]{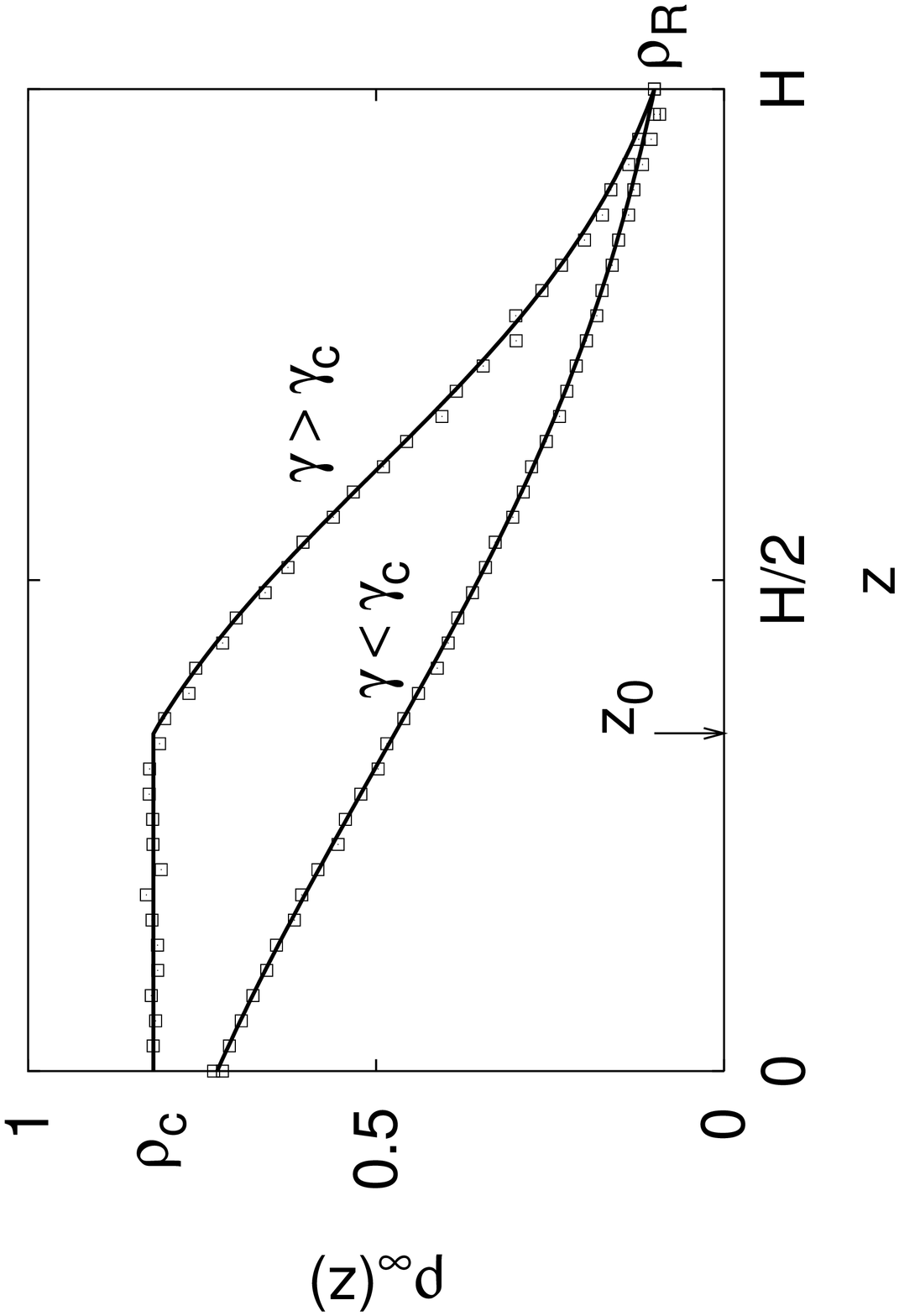}
\caption{Examples of the 
stationary profiles above and below the critical
temperature (solid lines). 
For $\gamma \le \gamma_{\rm c}(\rhor)$ (high temperature)
the equilibrium density profile is $\rho_\infty(z)=1/[\exp(\gamma z+\eta)+1]$,
with $\eta=\ln\left[(1-\rhor)/\rhor\right]-\gamma H$. 
At low temperatures, $\gamma  >\gamma_{\rm c}(\rhor)$, the stationary 
profile is $\rho_\infty(z)=\rho_{\rm c}$ for 
$z \le z_{\rm 0}(\rhor) \equiv 
H \left[1- \gamma_{\rm c}(\rhor)/\gamma \right]$; 
and $\rho_\infty(z)=1/[\exp(\gamma z+\eta)+1]$ for $z>z_{\rm 0}(\rhor)$.
The squares are the asymptotic densities obtained through simulation 
of the gravity-driven KA model on the BCC lattice (see~\cite{SeAr00} for
details on simulation).
}
\label{profile}
\end{figure}

The boundary conditions require vanishing of the current at $z=0$, 
$J(0,t)=0$; and $\rho(H,t)=\rhor$.
The stationary state is obtained when $\partial \rho/\partial t=0$, 
which implies that $J(z,\infty)=0$ for all $z$.
The jamming transition corresponds to the locus in the parameter space
$(\gamma,\, \rhor)$, at which a layer of critical density first appears,
$\rho(0,\infty)=\rho_c$.  This happens when 
\begin{eqnarray}
\gamma_{\rm c}(\rhor)  & = & 
  \frac{1}{H} \ln \left[ 
  \frac{ \rho_{\rm c} (1-\rhor) }{ 
  \rhor ( 1-\rho_{\rm c} ) } \right] \, .
\end{eqnarray}
Depending on the value of $\gamma$ two very different types of 
stationary profiles are 
found.  In Fig.~\ref{profile} we compare these with the stationary
profiles found from the Monte Carlo
simulation of the gravity-driven KA model on the BCC lattice~\cite{SeAr00}. 
A perfect agreement is obtained with no 
adjustable parameters. It is important to note that for
$\gamma>\gamma_{\rm c}(\rhor)$ the stationary profiles are no
longer equilibrium distributions, i.e. they do not
minimize the Helmholtz free energy functional Eq.~\ref{2}. 
%
%
%
%


\section{High-temperature phase, $\gamma<\gamma_{\rm c}(\rhor)$}

\begin{figure}[h]
\twofigures[width=5cm, angle=270]{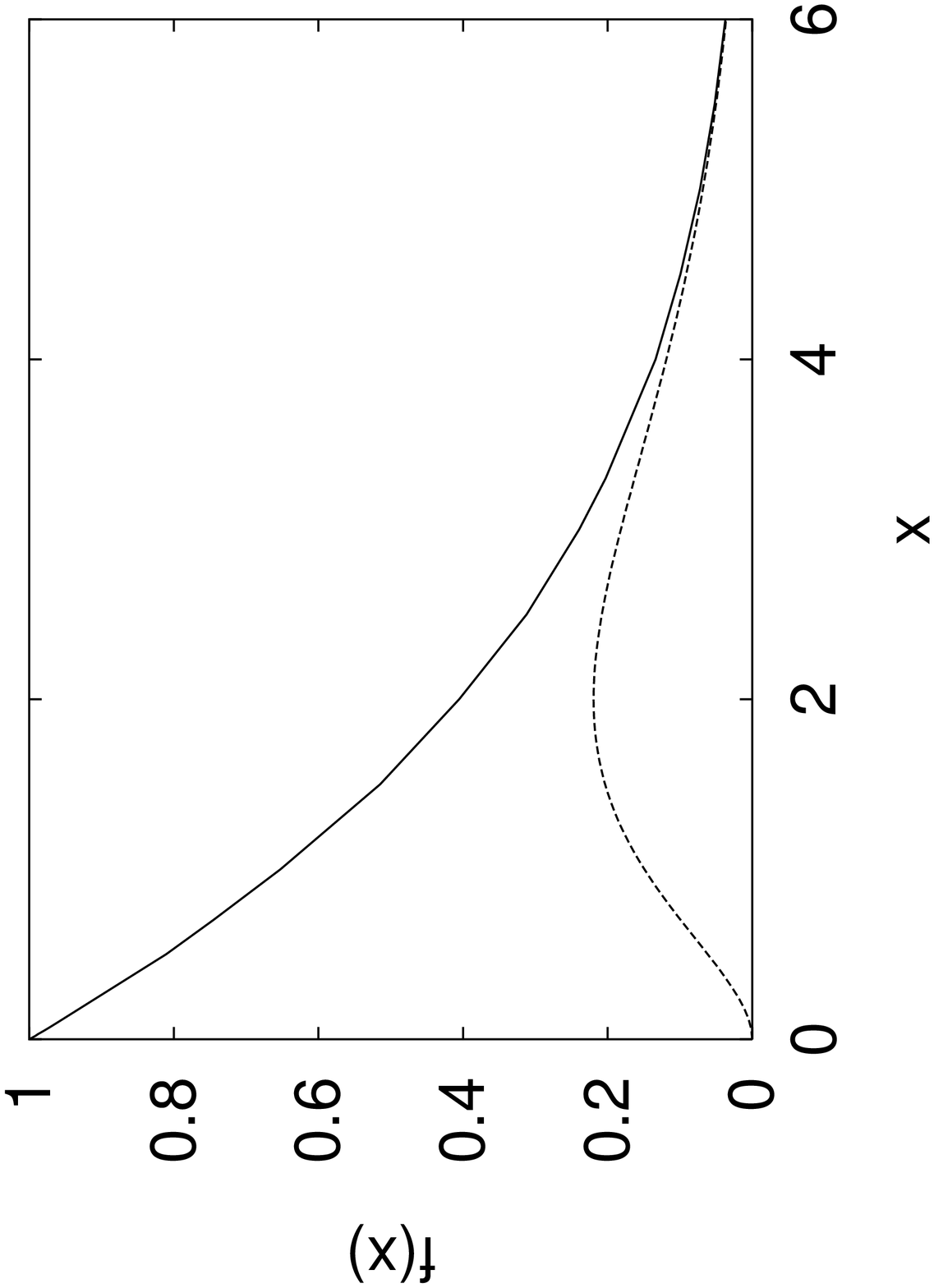}{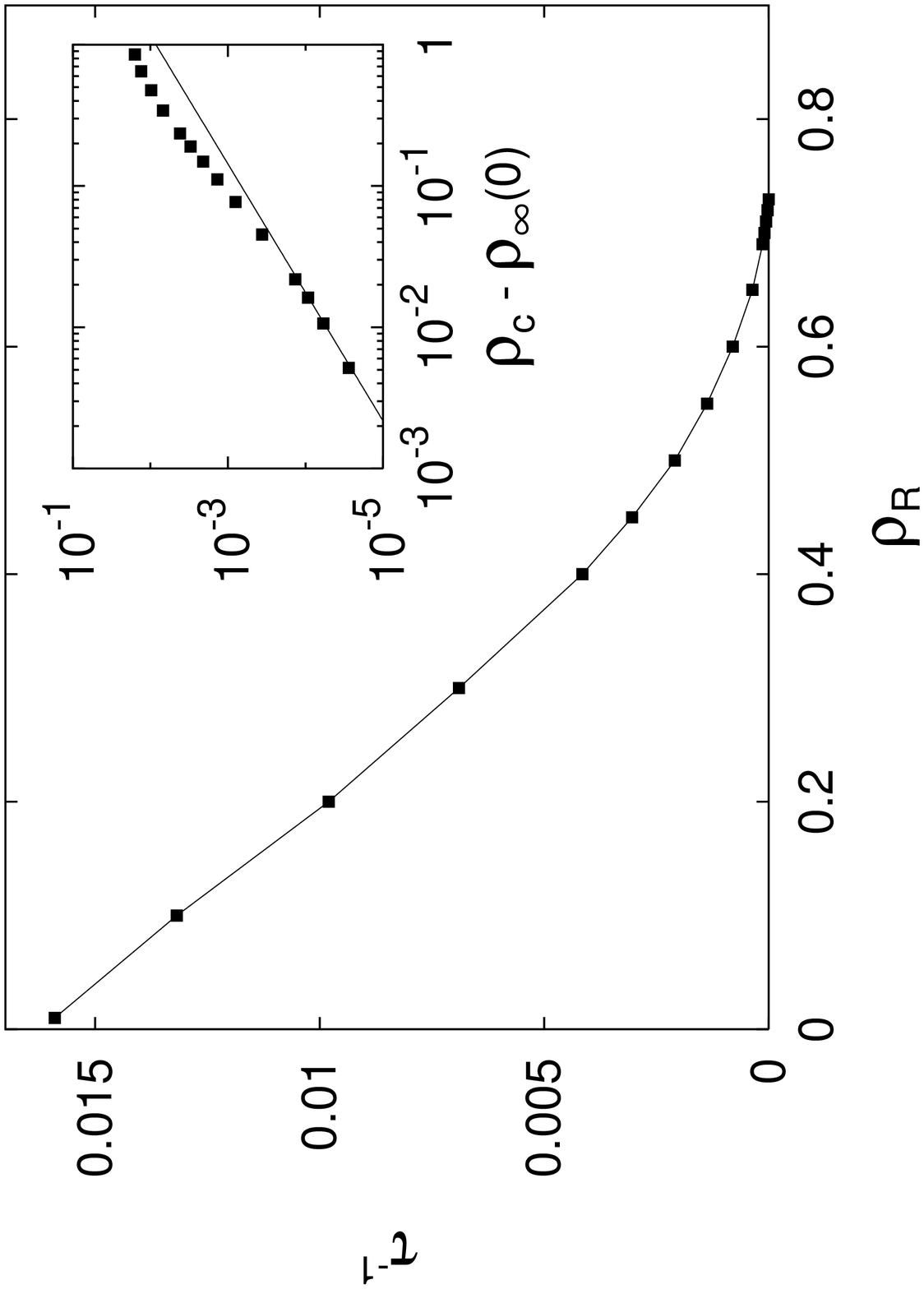}
\caption{Scaling function $f(x)$ (solid curve) and
 its asymptotic form (dashed line) as discussed in text.}
\label{Fig2}
\caption{Inverse relaxation time, $\tau^{-1}$, for $\gamma=0.1$ and $ 
H=10$ with $\rho_c=0.88$ and $\phi=3.1$, corresponding to the
Kob-Andersen model on a simple cubic lattice.
The points are the result of numerical integration of 
Eq.~(\ref{3}).  Inset shows the same data on the log scale. The 
characteristic time  diverges with exponent $\phi-2$
as the jamming transition is approached.}
\label{Fig4}
\end{figure}

At high temperature we expect an exponential relaxation 
towards the equilibrium 
distribution,
$\rho(z,t)\asymp \rho_\infty(z)+g(z){\rm e}^{-t/\tau}$.
We have solved Eq.~(\ref{3}) numerically and 
checked that the approach to equilibrium is indeed exponentially 
fast for any $\gamma<\gamma_{\rm c}(\rhor)$. 
To see how the characteristic time $\tau$ 
depends on the various parameters of the 
system we perform a scaling analysis of Eq.~(\ref{3}).
We first note that associated with the two right-hand terms of 
Eq.~(\ref{3}) there are two temporal scales.  
These can be identified as the diffusion time $\tau_1 \propto H^2$, 
and the drift time $\tau_2 \propto H/\gamma$. 
The characteristic time must, therefore, satisfy an {\it exact} 
equation $\tau^{-1}=\frac{\pi^2}{4 H^2} {\cal F}(\gamma H;\rhor)$,  
where ${\cal F}(x;\rhor)$ is a scaling function. 
The reason for the prefactor  $\pi^2/4$ will become clear from the forthcoming 
analysis. 
To further explore the properties of Eq.~(\ref{3}) we
now study its linear version, 
\begin{equation}
\label{6}
\frac{\partial \rho(z,t)}{\partial t}=
\frac{\partial^2 \rho}{\partial z^2} + 
\gamma \, \frac{\partial \rho}{\partial z} \,.
\end{equation}
For Eq.~(\ref{6}), the characteristic time of approach to 
equilibrium can be related to 
the singularities of the temporal Laplace transform of the density profile.
For $\tau>4/\gamma^2$ the characteristic time is found to satisfy
\begin{eqnarray}
\label{7}
\exp{\left\{H\sqrt{\gamma^2-4\tau^{-1}}\right\}}
=
\frac{\gamma+\sqrt{\gamma^2-4\tau^{-1}}}
{\gamma-\sqrt{\gamma^2-4\tau^{-1}}}\;,
\end{eqnarray}
while for  $\tau<4/\gamma^2$ it satisfies
\begin{eqnarray}
\label{8}
\cos{\left\{H\sqrt{4 \tau^{-1}-\gamma^2}\right\}}=
\tau \gamma^2/2-1 \,.
\end{eqnarray}
We note that for $\gamma=0$, $\tau^{-1}=\pi^2/4 H^2$. 
For finite $\gamma$ the inverse relaxation time for {\it linear} 
Eq.~(\ref{6}) can be written in a scaling form,
$\tau^{-1}=\frac{\pi^2}{4 H^2} f(\gamma H)$.  
Unlike the scaling function ${\cal F}(x;\rhor)$ for the {\it non-linear} 
Eq.~(\ref{3}), $f(x)$ does not depend on the reservoir density.
For $x=0$, $f(0)=1$; for large $x$, $f(x)\asymp (4/\pi^2)x^2 \exp(-x)$.
The graphs of $f(x)$  and of its asymptotic form are 
presented in Fig.~\ref{Fig2}.

%
%
%
%
%
%
With the insight gained from the study of Eq.~(\ref{6})
we can now  explore the full non-linear Eq.~(\ref{3}).
We first
consider the special case of zero gravity~\cite{PeSe}.
For $\gamma=0$ particles diffuse freely from the reservoir until
a uniform density profile $\rho_\infty(z)=\rhor$ is established.
At a low reservoir density $\rhor \ll \rho_{\rm c}$, 
we expect that the non-linearities of Eq.~(\ref{3}) should
be irrelevant and the relaxation time should reduce
to  $\tau^{-1}\approx \frac{\pi^2}{4 H^2}$. 
Linearizing Eq.(\ref{3}) around the equilibrium state
we find that the relaxation time   
for $\gamma=0$ is
\begin{equation}
\label{9}
\tau^{-1}=\frac{\pi^2}{4 H^2 (1-\rhor)} \left(1-\frac{\rhor}{\rho_{\rm c}}
\right)^{\phi} \,,
\end{equation}
or equivalently 
${\cal F}(0;\rhor)=(1-\rhor/\rho_{\rm c})^{\phi}/(1-\rhor)$.
Eq.~(\ref{9}) is in perfect agreement with the numerical integration
of  Eq.~(\ref{3}). 
As expected, the relaxation time diverges as 
$\rhor \rightarrow \rho_{\rm c}$. The exponent characterizing
this divergence is $\phi$.

Finally in the presence of a gravitational field we find,
by numerical integration of Eq.(\ref{3}),
that as  $\gamma \rightarrow \gamma_{\rm c}(\rhor)$,
the density of the first layer approaches $\rho_c$, $\rho_\infty(0) 
\rightarrow \rho_c$,  and
${\cal F} \sim  (\gamma_c-\gamma)^{\phi-2}$. 
The relaxation time diverges with exponent $\phi-2$, see Fig.{\ref{Fig4}},
implying that the dynamics is faster than in the zero gravity case.
Comparing with eq.~(\ref{9}), we see that the jamming transitions in the 
homogeneous and inhomogeneous systems, therefore, belong to distinct 
universality classes.


\section{Low-temperature dynamics, $\gamma>\gamma_{\rm c}(\rhor)$}

\begin{figure}[h]
\twofigures[width=5cm, angle=270]{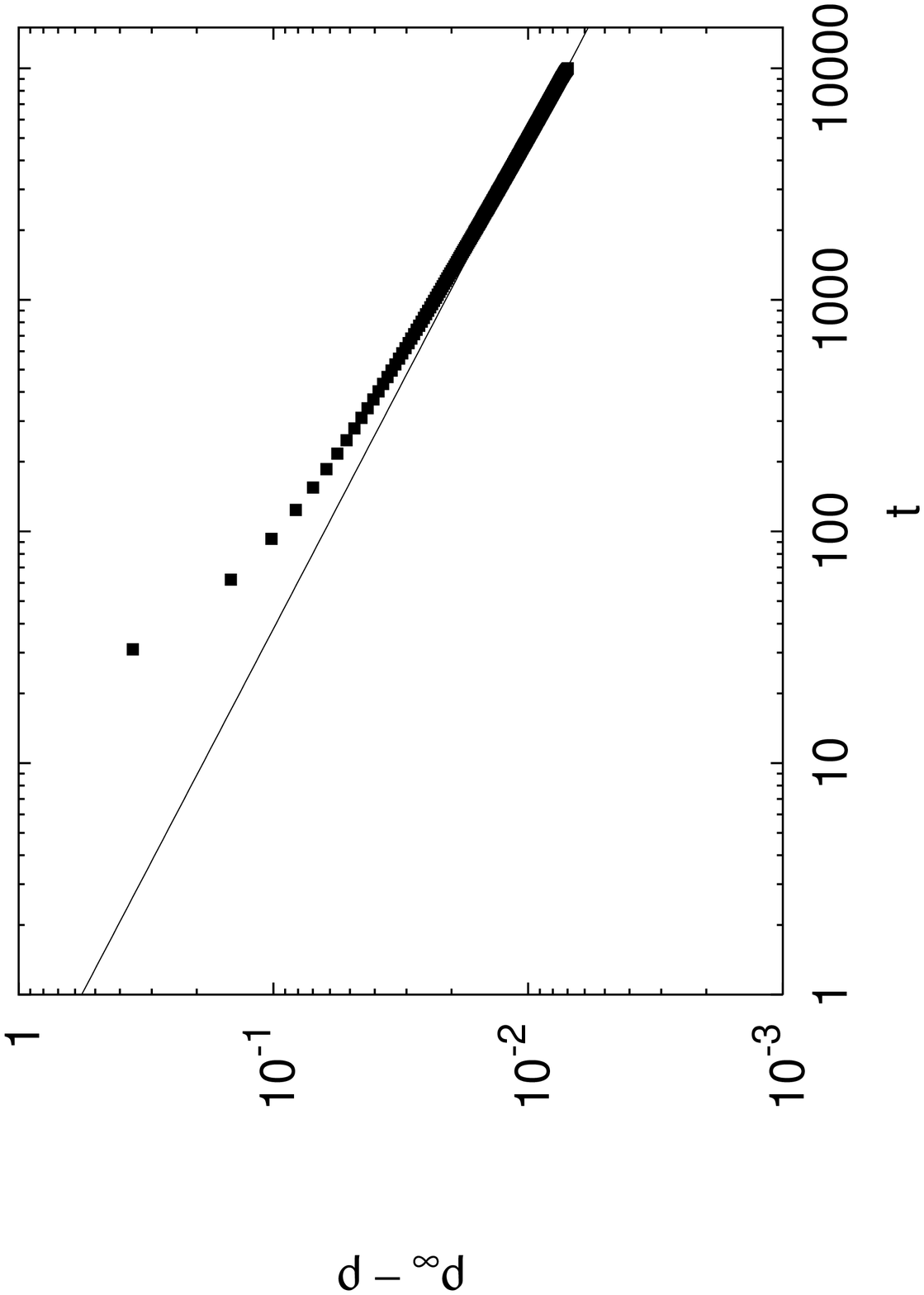}{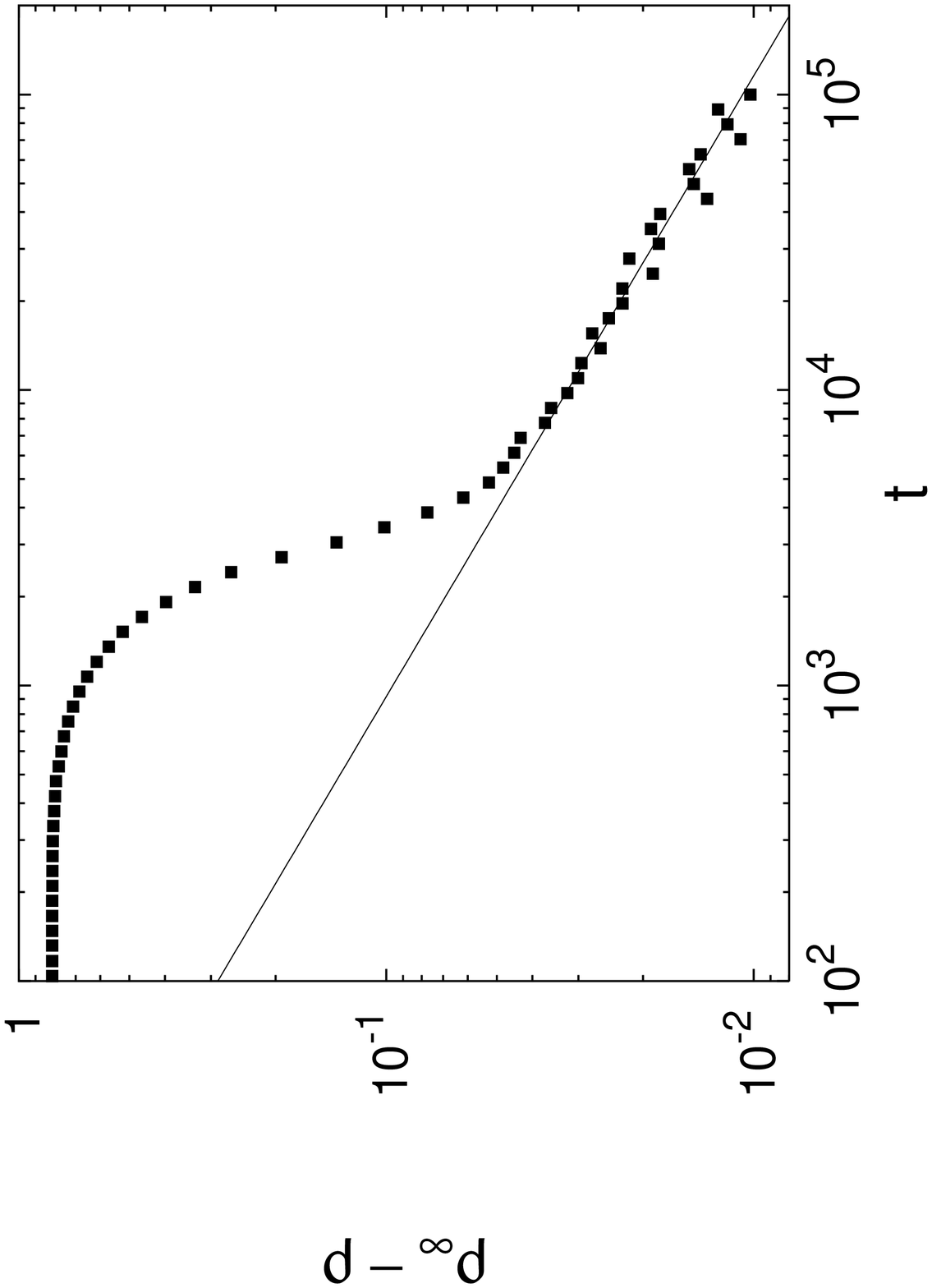}
\caption{Power-law density relaxation for
$\gamma > \gamma_{\rm c}(\rhor)$.
The  points are  result of  numerical integration of the 
Eq.~(\ref{3}), while the solid
line is the asymptotic solution given by Eq.~(\ref{5}).} 
\label{Fig1a}
\caption{Density relaxation
found in the  simulation of the 
gravity driven KA model on the BCC lattice. Averages are over 100 samples and
$H=4L=80$, $\rhor=0.1$, and $\gamma=0.1$. The line is the
theoretical prediction $t^{-1/(\phi-1)}$.}
\label{Fig1b}
\end{figure}

At low temperatures, the density of the lower layers closely
approaches the critical threshold, $\rho_c$.  
At such large packing fractions the movement of particles is 
highly restricted,
and even the slightest increase of density requires a rearrangement 
of a huge number of grains.
In this regime we expect slow relaxation and aging phenomena to appear. 
To check this supposition one should solve Eq.~(\ref{3}).
Unfortunately, due to its highly non-linear nature no analytic solution
is possible.  
Nevertheless an asymptotic solution can be found.  
Let us focus our attention on the bottom layers, $z<z_{\rm 0}(\rhor)$,
of the suspension, Fig.~\ref{profile}. 
At sufficiently long times, the density of these layers will be close to 
the critical one, $\rho(z,t) \simeq \rho_{\rm c}$.  
To the lowest order in 
$\Delta(z,t) \equiv 1-\rho(z,t)/\rho_{\rm c}$, 
Eq.~(\ref{3}) simplifies to,
\begin{equation}
\label{4}
\frac{\partial \Delta(z,t)}{\partial t}=
- \gamma\frac{\partial \Delta^\phi}{\partial z} \,.
\end{equation}
To solve this non-linear equation we propose a scaling ansatz
$\Delta(z,t)=\Delta(z/t^\alpha)$.
Substituting into Eq.~(\ref{4}) we see that this form is a solution 
if $\alpha=1$ and
\begin{equation}
\label{5}
\Delta(z,t)=\left[\frac{z}{ \phi\gamma\; t}\right]^\frac{1}{\phi-1} \,.
\end{equation}
To check the asymptotic solution, Eq.~(\ref{5}), we have numerically 
integrated Eq.~(\ref{3}) for $\gamma>\gamma_{\rm c}(\rhor)$.  
For large times a perfect agreement between the numerical solution  and
the Eq.~({\ref 5}) is found, as can be seen from Fig.~\ref{Fig1a}.
In the absence of gravity, $\gamma=0$, relaxation is slower and 
characterized by a different dynamic exponent, 
$ \Delta(z,t)  \sim t^{-1/\phi}$~\cite{PeSe}.
The same behavior is observed in the KA model 
(see ref.~\cite{PeSe} and Fig.~\ref{Fig1b}).

Up to now our discussion has been motivated by the dynamics of
the KA model which is characterized by a power law mobility coefficient. 
However, there are various systems for which  the mobility
vanishes according to the  Vogel-Fulcher law, $\Gamma(\rho)= 
\Gamma_0 \, \rho \, \exp \left[ \, a \,\rho_{\rm c}/
(\rho-\rho_{\rm c}) \right]$. 
In this case an analysis similar to the one presented above leads to
an asymptotic scaling solution, which for very large times reduces to
\begin{equation}
\label{21}
\Delta(z,t)=  \frac{a}{ \ln (t/z)} \,.
\end{equation}
%
Interestingly, a similar logarithmic law was discovered in the granular
compaction experiments~\cite{Knight95}, and is
supported by several analytical approaches~\cite{BoGe,Naim98,Brey,Head,Dean}.
%
%

We now turn to the discussion of the aging 
phenomena~\cite{Bouchaud,BoCuKuMe,sta.97,LaLe}.
Consider the two-times mean square displacement of grains, $B(t,t_{\rm w})$.
If $t$ is sufficiently larger than $t_{\rm w}$, this can be written as
\begin{equation}
\label{5_1}
  B(t,t_{\rm w})= \int_{t_{\rm w}}^t  
  ds \, \Gamma \left[ \rho(z,s) \right] \,.
\end{equation}
For power-law diffusion and  $z<z_{\rm 0}(\rhor)$  
we find, to leading order in $t$ and $t_{\rm w}$,
\begin{equation}
\label{5_2}
B(t,t_{\rm w}) \sim t_{\rm w}^{1-\mu} -  t^{1-\mu}  \,,
\label{B_a}
\end{equation}
with the exponent $\mu = \phi/(\phi-1)$.
Usually $\phi>1$, so that $\mu > 1$, which corresponds to a
super-aging regime~\cite{Bouchaud}.
This means that the effective structural relaxation 
time grows as $t_{\rm w}^{\mu}$,
what is faster than the age of
the system, $t_{\rm w}$~\cite{Bouchaud}.
This scaling behavior has been observed in the simulation
of the gravity-driven KA model~\cite{Mauro}. 
The comparison with the zero-gravity case, for which
a simple aging
($t/t_{\rm w}$) was found~\cite{PeSe}, 
once again shows the important role played by gravity.

For the Vogel-Fulcher law we find,  to the leading order
in $t$ and $t_{\rm w}$ (and $z<z_{\rm 0}(\rhor)$), a simple aging scenario,
$B(t,t_{\rm w}) \sim \log \left( t/t_{\rm w} \right)$.
It is interesting to observe the different behavior of this function
and~(\ref{B_a}) at finite waiting times 
$t_{\rm w}$.
In the former case,  
$\lim_{t \rightarrow \infty} B(t,t_{\rm w}) = \infty$, i.e., a weak
ergodicity scenario~\cite{Bouchaud92};
while in the latter, a finite limit is obtained
(which, however, vanishes as $t_{\rm w} \to \infty$).
Nevertheless, the manner in which 
time-translation invariance is violated is, in a sense, similar.
Indeed, if we consider the ``triangle relation''~\cite{CuKu},
$B(t_1,t_3) = f \left[ B(t_1,t_2) ,\, B(t_2,t_3) \right]$,
where the times $t_1,\, t_2$, and $t_3$ are in increasing order,
it is straightforward to check that in both Vogel-Fulcher and power
law cases
$f (x,y) = x + y $, implying that displacements over 
non-overlapping time intervals are statistically independent. 
This feature does not hold in the presence of activated aging 
for which $B(t,t_{\rm w}) \sim  \log t/\log t_{\rm w}$~\cite{BoCuKuMe}.

%


\section{Conclusions}

To summarize, we have presented an analytical study of 
the dynamics of a class of kinetic lattice models.
Our approach based on the local density approximation 
for the Onsager mobility allows us to predict  
the density relaxation law and the nature of aging behavior. 
The results reproduce successfully the behavior observed in the 
numerical simulation of the KA and 
gravity-driven KA model, and they could be relevant to systems 
for which steric hindrance and cage effect are dominant, such as
polydispersed colloidal suspension in gravitational
field~\cite{got92,tok95}, and slowly compacting dense granular matter.   
For hard-sphere systems it is known that the diffusion coefficient 
vanishes as a power-law 
with $\phi$ between $2$ and $3$, and
the sedimentation profile is approached very slowly (one year, 
roughly)~\cite{Chaikin}; 
hence a super-aging behavior in the two-times mean-square height 
displacement with exponent $\mu$ between $3/2$ and $2$ should be 
observed.


\stars
We thank J. Kurchan and D. Stariolo for discussions and a critical 
reading of the manuscript. We are grateful to the  referee 
for noticing  an error in Eq.~(\ref{9}) of the original manuscript.
This work was supported in part by the Brazilian agencies
CNPq and FAPERGS.
MS is supported by a Marie Curie fellowship of the EU
(contract ERBFMBICT983561).


\end{document}